\documentclass[aps,prb,final,twocolumn,showpacs]{revtex4}
\usepackage{graphicx}% Include figure files
\usepackage{dcolumn}% Align table columns on decimal point
\usepackage{bm}% bold math
\usepackage{color}

\begin{document}

\title{Optical Dichroism by Nonlinear Excitations in Graphene Nanoribbons}

\author{C. E. Cordeiro$^1$, A. Delfino$^1$,  T. Frederico$^2$, O. Oliveira$^{2,3}$ and
W. de Paula$^{4}$}

\affiliation{$^1$Instituto de F\'isica, Universidade Federal Fluminense, 24210-3400- Niter\'oi - RJ, Brazil \\
                 $^2$Departamento de F\'{\i}sica, Instituto Tecnol\'ogico de Aeron\'autica,
                               12.228-900, S\~ao Jos\'e dos Campos, SP, Brazil \\
                 $^3$Departamento de F\'{\i}sica, Universidade de Coimbra, 3004-516 Coimbra, Portugal \\
                 $^4$Departamento de F\'{\i}sica, Universidade Federal de S\~ao Carlos, 13565-905 S\~ao Carlos, SP,
                          Brazil}

\date{\today}

\begin{abstract}
The honeycomb carbon structure of graphene and nanotubes has a
dynamics which can give rise to a spectrum. This can be excited
via the interaction with an external electromagnetic field. In
this work, non-linear waves on graphene and nanotubes associated
with the carbon structure are investigated using a gauge model.
Typical energies are estimated and there scaling with the
nanoribbon width investigated. Furthermore, the soliton-photon
interaction depends on the incident photon polarization. In
particular, we find that the nanoribbon is transparent when the
polarization is along the largest length. Relying on the scaling
with the width, we suggest a way to experimentally identify the
soliton waves in nanoribbons.
\end{abstract}


 \maketitle

%====================================================================
%====================================================================
\section{Introduction and Motivation \label{introducao}}

The physics of nano-materials has been the subject of intense investigation in the past years.
In particular, graphene and nanotubes
\cite{wallace1947,novoselov2004,abergel2007,blake2007,casiraghi2007}
have been studied due to its potential technological applications and its peculiar electronic properties,
see \cite{castro2009,Peres2010}.

A popular approach to understand graphene and nanotubes is to use tight-binding or Hubbard like
models to describe the behavior of electrons and holes. Despite its success, these type of models do
not take into account the dynamics of the honeycomb array of carbon atoms. Therefore, dynamical effects
connected directly with the carbon structure are not taken into account by those type of models.

A chiral gauge model for graphene was suggested in \cite{Jackiw_Pi_2007}. In \cite{Oliveira2011} it
was studied its generalisation to include other gauge groups, not necessarily chiral invariant.
%In \cite{Jackiw_Pi_2007,Oliveira2011}  it was suggested a gauge
%model for graphene.
The models are able to accommodate the main
features of the electronic properties in graphene, like the
creation of mass gaps or quantum Hall effects, and parameterizes
the dynamics of the carbon background structure via a charged
scalar field $\varphi$ and
%a photon-like
spin one fields $A_\mu$.
The model allows to go beyond the electronic properties and to
investigate phenomena which are related with the dynamics of the
carbon crystal structure. Indeed, the authors explored the
non-linear dynamics of $\varphi$ and $A_\mu$ to compute mass gaps,
to derive finite energy vortex solutions,
which can lead to flux quantisation and Aharonov-Bohm effects (see e.g. \cite{JackiwPRB2009} and
references therein)
and to check that the
gauge theory is compatible with the experimentally observed
electronic quantum Hall effect.
Furthermore, the fermionic zero modes associated to the finite energy vortex configurations
 \cite{Oliveira2011}  are solutions of a one dimensional Schr\"odinger-like equation with a scale invariant $1/z^2$
 potential \cite{Jackiw1972}. Certainly, the theory can accommodate other phenomena not explored so far and, in
particular, phenomena whose dynamics is linked with the honeycomb
array of carbon atoms.

The complex scalar field $\varphi$ and $A_\mu$ model the dynamics of the carbon crystal structure,
on top of which electrons and holes live. Certainly, the electronic properties of graphene are sensitive to the
$\varphi$ and $A_\mu$ configurations. Changing the configuration of the bosonic fields, the electrical
and thermal transports properties associated with the fermionic degrees of freedom change accordingly.
For example, phonons are associated with fluctuations of $\varphi$ and $A_\mu$ around a given configuration
and contribute to the graphene specific heat. Then, the investigation of the thermal properties of graphene can help
understanding the bosonic part of the gauge model. On the other hand, the scalar field $\varphi$ being
a charged field, it couples directly to the electromagnetic field $C_\mu$ and can contribute to the graphene
optical properties. Therefore, within the gauge model of \cite{Jackiw_Pi_2007,Oliveira2011}, it is of paramount importance to understand
how the bosonic degrees of freedom can be excited via the coupling with an external field.
If one is able to control the $\varphi$ and $A_\mu$ configurations by coupling with external fields,
one can achieve a better control of graphene and, in general, nano-materials properties.

From the pure field theory point of view, $\varphi$ couples directly to the fermions and, for example,
can trigger the generation of a fermionic mass gap. Given that $\varphi$ is a dynamical field, the model
allows us to define regions where the mass gap vanishes, while in other regions the fermions acquire a
finite mass with the corresponding decreasing of its mobility, i.e. from the point of view of the gauge model
the mass gap is, itself, a dynamical quantity. One would like to understand what type of $\varphi$ configurations
are allowed and how they impact on the electronic properties.

The self-couplings of $\varphi$ allow for solitonic solutions of
the fields equations. In this work, we will use the term solitonic
solution to describe any solution of the non-linear differential
equation of motion for $\varphi$. The field fluctuations around
the soliton can give rise to phonons with new types of dispersion
relations. One expects contributions to the optical and thermal
properties of graphene coming from the solitonic waves.
Furthermore, $\varphi$ has a non-vanishing electric charge and the
non-linear waves couples with an external electromagnetic field.
This coupling opens the possibility of generating these non-linear
modes by exciting graphene with a laser or a properly tuned
electromagnetic field.

The gauge model suggested in \cite{Oliveira2011} is an effective quantum field theory for the dynamics of a
2+1 dimensional system. In principle, it can be applied to any 2D nano-materials, other than graphene.
In this work we will also consider nanotubes. From the point of view of the gauge theory, the main difference
between the two systems is the geometry of the two carbon based materials. If graphene is associated with a 2D
flat sheet, a nanotube is a two dimensional material having the same type of unit cell but with a cylindrical geometry.

In this paper, we investigate the classical solutions of the
non-linear bosonic equations of motion of the gauge model
\cite{Oliveira2011} for graphene and nanotubes. Besides computing
explicitly non-linear configurations for $\varphi$ and $A_\mu$, we
also discuss how these nonlinear modes couple with an external
electromagnetic field. In particular, we find that the coupling
between the nonlinear modes and the photon field displays
dichroism, i.e. the ability to absorb light depends on the
polarization state of the incoming photon. Dichroism with a strong
absorption anisotropy from the electronic intraband transitions
requiring a finite chemical potential was investigated in
\cite{Hipolito2012}. For these two cases, the pattern of the
coupling with the polarization of the photon is similar, i.e. the
graphene nanoribbon is transparent when the polarization is along
the largest length of the carbon material, but also the energy
range where dichroism is expected to be seen is in the far
infrared and terahertz frequencies. If the intraband transition
requires a non-vanishing chemical potential $\mu$ to generate
dichroism, the coupling with the nonlinear waves is independent of
$\mu$. This observation enables to experimentally distinguish
between the two mechanisms.

The paper is organized as follows. In section \ref{the_model} we review the gauge model. The coupling
with an external electromagnetic field is discussed, the equations of motion and the Hamiltonian for
 the gauge model are
derived. In section \ref{classical_solutions} the solitonic
solutions of the classical equations are discussed in the present
context. In section \ref{electro} we investigate the
soliton-photon transition amplitude and decay rate. Finally, in
section \ref{fim} we resume and conclude.

%====================================================================
%====================================================================
\section{Gauge Model for Graphene and Nanotubes \label{the_model}}

For completeness, in this section we resume the main features of the gauge model described in \cite{Oliveira2011}.
The interested reader should look at the cited paper for a detailed discussion.

The model describes electrons and holes via a Dirac equation in two dimensions using a four component spinor
\begin{equation}
 \Psi ~ =  ~ \left( \begin{array}{c}
\psi^b_+ \\ \psi^a_+ \\ \psi^a_- \\ \psi^b_- \end{array} \right) \, ,
\end{equation}
where the indices $a$ and $b$ refer to the two triangular sublattices of graphene and $+(-)$ to the two Dirac
$K (K^\prime)$ points.
Throughout this paper, we will use the following representation for the Dirac matrices
\begin{equation}
 \gamma^0 = \left( \begin{array}{ll} 0 & 1 \\  1 & 0 \end{array} \right), ~
 \vec{\gamma} = \left( \begin{array}{ll} 0 & \vec{\sigma}  \\  - \vec{\sigma} & 0 \end{array} \right),
  ~
 \gamma_5 = \left( \begin{array}{ll} -1 & 0  \\  0 & 1 \end{array} \right);
\end{equation}
$\sigma^j$ stand for the Pauli matrices.

The dynamics of the carbon background structure is associated with a charged scalar field $\varphi$
and with a vector field $A_\mu$. The vacuum expectation value of the scalar field $\langle \varphi \rangle$
vanishes for pure graphene, while for doped or deformed graphene $\langle \varphi \rangle \ne 0$.
If $\langle \varphi \rangle \ne 0$, the model generates a fermion mass via spontaneous symmetry breaking.

The lagrangian density for the gauge model reads
\begin{eqnarray}
  \mathcal{L}  & =  & \overline\psi \, i \, \gamma^\mu D_\mu \psi
   -    g_2 \, \left( \varphi^\dagger \varphi \right) \, \overline\psi \psi
    -  i \, h_2 \, \left( \varphi^\dagger \varphi \right)\, \overline \psi \gamma_5 \psi
                        \nonumber \\
  & & \qquad                        + ~
   D^\mu \varphi^\dagger D_\mu \varphi \, - \, V( \varphi^\dagger \varphi ) \nonumber \\
   & & \qquad  - \, \frac{1}{4} F^{\mu\nu}F_{\mu\nu},
  \label{new_lagrangian}
\end{eqnarray}
where the fermionic covariant derivative is $D_\mu = \partial_\mu
+ i g A_\mu$, with a similar expression for the bosonic covariant
derivative after replacing $g$ by $g_\varphi$,
\begin{equation}
   V(\varphi^\dagger \varphi) ~ = ~ \mu^2 \left( \varphi^\dagger\varphi \right) ~ + ~ \frac{\lambda_4}{2} \left( \varphi^\dagger\varphi \right)^2
                          + ~ \frac{\lambda_6}{3} \left( \varphi^\dagger\varphi \right)^3
  \label{potential_phi}
\end{equation}
up to a constant $V_0$, and
\begin{equation}
  F_{\mu\nu} = \partial_\mu A_\nu - \partial_\nu A_\mu \,
\end{equation}
is the usual Maxwell tensor for the vector bosonic field.

In what concerns the electrons and its interactions with $\varphi$ and $A_\mu$, the lagrangian (\ref{new_lagrangian}) is a
particular case of the possible theories studied in \cite{Oliveira2011}. In the following, we will consider only the solitonic
solutions of the scalar field equation of motion.
The lagrangian (\ref{new_lagrangian}) is an example of fermionic dynamics.

In (\ref{potential_phi}), the coupling constants $\mu^2$, $\lambda_4$, $\lambda_6$ parametrize the self-couplings of the carbon structure
and are associated with the carbon mechanical properties like phonon masses, phonon self-interactions and graphene stiffness.

The lagrangian density (\ref{new_lagrangian}) is invariant under
local $U_f(1) \otimes U_b(1)$ transformations. Recall that
$\varphi^\dagger \varphi$ can be identified with a dynamical
fermionic mass that is a function of space and time. Given that
for doped graphene $\langle \varphi \rangle \ne 0$, this term
generates a fermionic mass gap, which vanishes for the ground
state of pure graphene. Besides the generating of a mass term for
the fermions, the model also gives masses for the scalar and
vector degrees of freedom. The model allows for vortex solutions
for the pure bosonic sector of the theory.

One of the goals of the present work is to investigate how the bosonic degrees of freedom in (\ref{new_lagrangian})
can be excited via the coupling with an external electromagnetic field $C_\mu$. The coupling with
$C_\mu$ is introduced in the usual way, i.e. via the minimal prescription, where a $i  e \, C_\mu$ should be added
to the covariant derivative, where $e$ stands for a generic electric charge.

%=======================================================================
%=======================================================================
\subsection{Equations of Motion}

The equations of motion associated with the various fields are derived from $\mathcal{L}$ in the usual way.
For fermions, they are given by
\begin{equation}
  \Big\{ i \gamma^\mu D_\mu   -   \left( g_2 + i h_2 \, \gamma_5 \right) \left( \varphi^\dagger \varphi \right)
                                                  \Big\} \psi~ = ~ 0 \, ,
  \label{equation_fermion}
\end{equation}
where $D_\mu = \partial_\mu + i g A_\mu + i e \, C_\mu$ is the covariant derivative.
The corresponding equation for the scalar field $\varphi$ is
\begin{eqnarray}
 D^\mu D_\mu \varphi  & = &    - \, g_2\, \varphi \, \overline\psi \,\psi   - i h_2 \, \varphi  \, \overline\psi \, \gamma_5\psi
  \nonumber \\
  &  & - \Big\{ \left( \mu^2 + e^2_\varphi  \left( C^z \right)^2  \right)    \nonumber \\
  &  & \qquad\qquad\quad     +  \lambda_ 4 \left( \varphi^\dagger \varphi \right)
                                                +   \lambda_6 \left( \varphi^\dagger \varphi \right)^2 \Big\} \varphi
  \label{equation_boson}
\end{eqnarray}
with the covariant derivative being $D_\mu = \partial_\mu + i g_\varphi A_\mu + ie_\varphi C_\mu$ and
$e_\varphi$ being the effective electric charge of $\varphi$.
The gauge field equation of motion reads
\begin{equation}
   \partial_\mu F^{\nu\mu} =
   - g \, \overline\psi \gamma^\nu \psi - i \, g_\varphi \, \varphi^\dagger \left( D^\nu \varphi \right)
                           + i \, g_\varphi  \left( D^\nu \varphi \right)^\dagger \varphi \, .
    \label{equation_photon}
\end{equation}

In what concerns the coupling with the external electromagnetic field $C_\mu$, the lagrangian density and
the equations of motion show that, to lowest order in perturbation theory, the coupling with $\varphi$ is of order
$e_\varphi$,  while $A_\mu$ is of the order $g_\varphi e_\varphi$ or $e^2_\varphi$.
Then, one expects the degrees of freedom associated with $\varphi$ to have a high probability of being excited
through the incidence of an electromagnetic wave.

%=========================================================================
%=========================================================================
\subsection{Hamiltonian Density \label{hamiltoniano}}

The Hamiltonian density $\mathcal{H}$ can be read from the energy-momentum tensor
\begin{equation}
 T^{\mu\nu} ~ = ~
 \sum_{ u = \overline\psi , \psi, \varphi^\dagger, \varphi, A_\mu}
 \frac{\partial \mathcal{L}}{\partial \left( \partial_\mu u \right)} \, \partial^\nu u ~ - ~ g^{\mu\nu} \mathcal{L} \, .
\end{equation}
It follows that
\begin{widetext}
\begin{eqnarray}
  \mathcal{H}  =  T^{00} & = &
   - i  \overline\psi  \vec{\gamma} \cdot \left( \nabla - i g \vec{A} - i e \vec{C} \right) \psi +
   \overline\psi  \gamma^0 \big( g A^0 + e C^0 \big) \psi
%   \nonumber \\
%   & &
%   \qquad
               + ~ g_2 \left(\varphi^\dagger\varphi\right) \, \overline\psi \, \psi
               +  i h_2 \left(\varphi^\dagger\varphi\right) \, \overline\psi \,\gamma_5 \psi
%\nonumber \\
%   & &
%   \qquad
%               - ~ e \, C^z \overline\psi \,\gamma^z \psi
               \nonumber \\
  & &
   ~ +~ \pi^\dagger \, \pi  ~ + ~ \vec{D} \varphi^\dagger \cdot \vec{D} \varphi  + V \left( \varphi^\dagger \varphi \right)
%   \nonumber \\
%  &  &
%   \qquad
%+ ~e^2_\varphi \left( C^z  \right)^2 \left(\varphi^\dagger\varphi\right)
%\nonumber \\
  ~ + ~ \frac{1}{2} \left( \vec{E}^2 + \vec{B}^2 \right)
  \label{eq_hamiltoniano}
\end{eqnarray}
\end{widetext}
where $\pi^\dagger = \partial^0\varphi$ and $\pi = \partial^0\varphi^\dagger$ are the canonical momentum density
fields associate with $\varphi$ and $\varphi^\dagger$, respectively.
The vectors $\vec{E}$ and $\vec{B}$ are the usual electric and magnetic fields associated with the
gauge vector $A_\mu$.

%=========================================================================
%=========================================================================
\section{Soliton-like Solutions for Graphene and Nanotubes \label{classical_solutions}}

In this section we will discuss the solutions of classical equations of motion.
Let us for the moment disregard the contributions coming from fermions and gauge fields.
The model assumes that the scalar and vector fields are associated with the carbon lattice background.
The Dirac electrons are soft degrees of freedom and one expects that the fermion-scalar couplings are subleading
for the dynamics of the scalar field.

Assuming that the only dynamical field is $\varphi$, the corresponding classical configurations
are obtained solving  equation (\ref{equation_boson}).
In order to compute solitonic like solutions, we will consider the following ansatz
\begin{equation}
  \varphi ( \vec{r} , t ) = \phi(x - v t) \, e^{iky}
  \label{graphene_ansatz}
\end{equation}
for a graphene sheet living on the $xy$--plane and
\begin{equation}
  \varphi ( \vec{r} , t ) = \phi(z - v t) \, e^{i n \theta} \, ,
  \label{nanotube_ansatz}
\end{equation}
for a nanotube whose symmetry axis is the $z$-axis. In order to describe the nanotube we use cylindrical coordinates
$r$, $\theta$ and $z$.

The ansatze (\ref{graphene_ansatz}) and (\ref{nanotube_ansatz}) reduce the partial
differential equation (\ref{equation_boson})  to an ordinary differential equation in the variable $\zeta$, where
\begin{equation}
  \zeta = \left\{ \begin{array}{lll}
         x - v t, & & \mbox{ for graphene,} \\
         & & \\
         z - v t, & & \mbox{ for a nanotube.}
         \end{array} \right.
\end{equation}
In terms of $\zeta$, equation (\ref{equation_boson}) becomes
\begin{equation}
 (1 - v^2) \phi^{\prime\prime} = \left( \mu^2 + k^2 \right) \phi + \lambda_4 \phi^3 + \lambda_6 \phi^5 \, ,
 \label{equation_varphi_solitonic}
\end{equation}
where a prime means a derivative with respect to $\zeta$.

The ansatz (\ref{nanotube_ansatz}) used to describe nanotubes, makes the equation of motion
equal to (\ref{equation_varphi_solitonic}) after the replacement of $k^2$ by $n^2/R^2$.
In the following, we will explore the solutions of equation (\ref{equation_varphi_solitonic}) for graphene,
but one should have in mind that the results for graphene have a directly translation into nanotubes.

The energy associated with the solitonic waves (\ref{graphene_ansatz}) is given by
\begin{widetext}
\begin{equation}
E_{graph} ~ = ~ W \int^{L_x/2}_{-L_x/2} dx ~ \Big\{ (1 + v^2) \left(\phi^\prime\right)^2
                                                            \, + \, \left( \mu^2 + k^2 \right) \phi^2 \, + \, \frac{1}{2} \lambda_4 \phi^4
                                                            \, + \, \frac{1}{3}\lambda_6 \phi^6 \Big\} \, ,
\label{E_graph}
\end{equation}
\end{widetext}
where $L_x$ and $W$ are the dimensions of the graphene sheet. For a nanotube of radius $R$ and length $L$,
the energy associated with the non-linear wave (\ref{nanotube_ansatz}) is given by (\ref{E_graph}) after the
replacement of $W$ by $2 \pi R$ and of $k^2$ by $n^2/R^2$.
%\begin{widetext}
%\begin{equation}
%E_{nano} ~ = ~ 2 \, \pi \, R \,  \int^{L_z}_{-L_z} dz ~ \Big\{ (1 + v^2) \left(\phi^\prime\right)^2
%                                                            \, + \, \left( \mu^2 + \frac{n^2}{R^2} \right) \phi^2 \, + \, \frac{1}{2} \lambda_4 \phi^4
%                                                            \, + \, \frac{1}{3}\lambda_6 \phi^6 \Big\} \, .
%                                                            \label{E_nano}
%\end{equation}
%\end{widetext}
For graphene and nanotubes
%From (\ref{E_graph}) and (\ref{E_nano}) it follows that
the lowest energy state is associate with $v = 0$, i.e.
for a static solution, and with $k = 0$ for graphene or $n = 0$ for nanotubes. Furthermore, for the static solution, i.e. when $v = 0$, minimizing the energy is equivalent to solve the classical equation motion
(\ref{equation_varphi_solitonic}).
% for both systems.

From the point of view of the differential equation
(\ref{equation_varphi_solitonic}), the solutions describing
travelling solitonic waves, i.e. for solutions with $v \ne 0$, are
the static solutions after the rescaling of the potential
parameters $\mu^2 + k^2$, $\lambda_4$ and $\lambda_6$ by the
inverse of $1 - v^2$. Note that for $v < v_f \approx c/ 300$ the
rescaling does not change the nature of the solution derived for
$v = 0$. However, for $v > v_f$ the rescaling formally changes the
sign of coupling constants and can lead to a change on the
functional form of the solution. Given that we aim to investigate
non-linear waves associated with the carbon structure of graphene
and nanotubes, one expects to be closer to the first case, $v <
v_f$, and we will not explore the situation where the solitonic
wave moves faster than the fermions $v > v_f$. From now on, we
will assume that $v = 0$ except where stated explicitly.

Before starting to discuss the solutions of the equation of motion (\ref{equation_varphi_solitonic}), let us define the notation.
In the following, the $k^2$ term will be included in the definition of the potential, i.e. we will write
\begin{equation}
 V(\phi) = \left( \mu^2 + k^2 \right) \phi^2 \, + \, \frac{1}{2} \lambda_4 \phi^4
                                                            \, + \, \frac{1}{3}\lambda_6 \phi^6 \, .
\end{equation}
Then, the energy becomes
\begin{eqnarray}
E & = & W \int^{L_x/2}_{-L_x/2} dx ~ \Big\{ \left(\phi^\prime\right)^2  + V( \phi ) \Big\} \nonumber \\
   & = & 2 \, W \int^{L_x/2}_{-L_x/2} dx ~  V( \phi ) \, ,
   \label{E_Vphi}
\end{eqnarray}
where to write the last expression we have used the equation of motion.

%====================================================================
%====================================================================
\subsubsection{Kink Solution \label{sec_kink}}

An analytical solution for (\ref{equation_varphi_solitonic}) can be obtained when $\lambda_6 = 0$ and
\begin{equation}
  V(\phi) = \frac{\lambda_4}{2}  \left( \phi^2 - \phi^2_0 \right)^2 \, .
\end{equation}
Note that a constant $V_0 = \lambda_4 \phi^4_2 /2$ was added to the original potential $V(\phi)$  in order to have a positive defined potential energy. In terms of the original potential one has $\lambda_6 = 0$ and $\mu^2 + k^2 = - \lambda_4 \phi^2_0$.
The classical configurations are
\begin{equation}
\phi (\zeta) = \pm \, \phi_0 ~\tanh \left\{ \phi_0 \sqrt{\frac{\lambda_4}{2}} \, ( \zeta - \zeta_0 ) \right\} \, ,
\label{kink}
\end{equation}
where $\zeta_0$ is a constant of integration. The plus (minus) sign solution is known in the literature
as kink (anti-kink).  For an infinite length sheet of graphene, the energy associated with (\ref{kink}) is given by
\begin{equation}
E_{sol} =  \frac{4}{3} \, \sqrt{2 \lambda_4} \, \phi^3_0 \, W \, .
\label{E_kink}
\end{equation}

%====================================================================
%====================================================================
\subsubsection{A Solution for a $\phi^6$ Potential \label{sec_loehe}}

The classical configuration associated with the potential energy
\begin{equation}
   V(\phi) = \frac{\lambda^2}{2}  \phi^2 \left( \phi^2 - \phi^2_0 \right)^2 \, ,
   \quad \lambda > 0
\end{equation}
was investigated by Lohe in \cite{Lohe1979}. The constants in the original potential are given by
$\lambda_6 = 3 \lambda^2/2$, $\lambda_4 = - 2 \lambda^2 \phi^2_0$ and $\mu^2 + k^2 = \lambda^2 \phi^4_0 /2$.
The equations of motion have the solutions
\begin{equation}
  \phi ( \zeta ) = \pm \frac{\phi_0}{\sqrt{2}} ~ \sqrt{ 1 \pm \tanh \left[ \frac{\lambda \, \phi^2_0}{\sqrt{2}} \, ( \zeta - \zeta_0 ) \right]  } \, .
  \label{lohe}
\end{equation}
The energy associated with any of these configurations  is given by
\begin{equation}
  E_{sol} = \frac{\sqrt{2}}{8} \, \lambda^2 \, \phi^4_0 \, W \,
  \label{energia_loehe}
\end{equation}
for a graphene with infinite length and for any combination of plus and minus signs.

\begin{figure}[t] %  figure placement: here, top, bottom, or page
   \centering
   \includegraphics[width=3in]{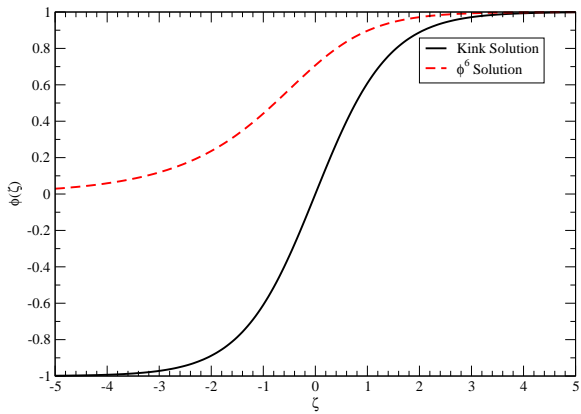}
   \caption{The solutions (\ref{kink}) and (\ref{lohe}) in arbitrary units. All constants where set to one and $\zeta_0 = 0$.}
   \label{fig:solucoes}
\end{figure}

The profile of solutions (\ref{kink}) and (\ref{lohe}) can be seen in figure \ref{fig:solucoes}.

%===========================================================================================
%===========================================================================================
\subsection{Energy Estimation \label{energy_estimation}}

The system of units used so far take $\hbar =1 $ and the Fermi velocity $v_F = 1$.
In order to convert to the usual system of units, we will use for the speed of light $c = 2.998 \times 10^8$ m/s,
$v_F = c / 300$ and $\hbar v_F = 0.6578$ eV nm. To estimate the energy associated with the solitonic
configuration, recall that in the gauge model the fermion gap is twice the fermion mass, i.e.
\begin{equation}
   E_g = 2 \sqrt{g^2_2 + h^2_2} ~ \phi^2_0 \,;
   \label{Eq_gap_E}
\end{equation}
see \cite{Oliveira2011} for details.
For nanoribbons, a typical value for the gap\cite{Li2008} being $0.1$ eV for $W = 10$ nm. For
$\sqrt{g^2_2 + h^2_2} \approx 1$, it follows that
\begin{equation}
    Ê\phi^2_0 = 0.05 ~\mbox{ eV}
\end{equation}
and
\begin{equation}
   E_{sol} ~ =  ~ \left( 0.23 ~\mbox{eV}^{1/2} \right) ~ \sqrt{\lambda_4} ~ \left(Ê\frac{W}{10 \mbox{ nm}} \right) \, ,
   \label{eq_energia_kink}
\end{equation}
 for the kink solution discussed in section \ref{sec_kink}, where $\lambda_4$ is given in eV and the width $W$ in nm,
and
\begin{equation}
    E_{sol} ~ =  ~ \Big( 6.7 ~\mbox{ meV} \Big) ~ \lambda^2 ~ \left(Ê\frac{W}{10 \mbox{ nm}} \right)
    \label{eq_energia_loehe}
\end{equation}
for the soliton of section \ref{sec_loehe}; recall that $\lambda$
is dimensionless. Note that according to (\ref{E_kink})  and
(\ref{energia_loehe}) (\ref{Eq_gap_E}), the energy associated with
the solitonic configurations is driven by the mass gap.
Furthermore, from (\ref{kink}) and (\ref{lohe}) one can estimate
the dimension $l$ of the transition region associated with each of
the solitonic solution. Indeed, from $\phi^2_0 = 0.05$ eV it
follows for the kink solution $l \approx 2 / \phi_0 \,
\sqrt{\lambda_4} =  14 / \sqrt{\lambda_4}$ nm, for $\lambda_4$
given in eV. For the other nonlinear wave $l \approx \sqrt{2} /
\lambda \, \phi^2_0 = 43 / \lambda$ nm.

In order to understand how the energy of the non-linear wave changes with $W$, one has to consider $E_g (W)$.
In \cite{Han2007} $E_g (W)$ was investigated experimentally and the authors conclude that for $W > 16$ nm,
the observed gaps are well described by
\begin{equation}
    E_g (W) = \frac{ \alpha }{W - W^*} \, ,
\end{equation}
where $\alpha = 0.2$ eV nm and $W^* = 16$ nm. Taking into account $E_g (W)$, it follows that the energy of the
solitonic solutions behave as
\begin{equation}
  E ~~ \propto ~~  \frac{W}{\left( W - W^* \right)^{3/2}} ~~ \mbox{ and } ~~ \frac{W}{ \left( W - W^* \right)^2 } \, ,
  \label{eq_e_width}
\end{equation}
where the first result is for the kink solution (see section \ref{sec_kink}) and the later for the solutions of section \ref{sec_loehe}.
In both cases $E$ vanish in the limit of infinite width, i.e. when graphene is recovered.
This result opens the possibility of tuning the energy of the nonlinear modes by chosen the nanoribbon width.

From the point of view of exciting these nonlinear modes of the carbon honeycomb structure, the results summarized
in equation (\ref{eq_e_width}) mean that as graphene becomes wider, it will be to easier to excite the solitonic modes.
In particular for bulk graphene these solitonic modes have zero energy and they must be a component of the
graphene ground state wave function.

As a side remark, we would like to point out that as the width becomes sufficiently small, the edges of the carbon
structure, being zig-zag or armchair-shapped, should be taken in the computation of the solitonic energy.

%===========================================================================================
%===========================================================================================
\section{Coupling Electromagnetic and the Nonlinear Waves \label{electro}}

The soliton modes are associated with a complex scalar field $\varphi$ and, therefore, they couple with the
electromagnetic field $C_\mu$ through and effective electric charge $e_{sol}$.
The Hamiltonian describing the interaction between $\varphi$ and $C_\mu$ is given by
\begin{eqnarray}
   H_I & = \int \, dx \,dy ~\Big\{ & ie_{sol} \, C^\mu \left[ \left(\partial_\mu \varphi^\dagger\right) \varphi -
                                                                       \varphi^\dagger \left( \partial_\mu\varphi\right) \right]  \nonumber \\
                                             & & \qquad + ~ e^2_{sol} \, C_\mu C^\mu \, \left( \varphi^\dagger\varphi \right) \Big\} \, .
\end{eqnarray}

For a classical scalar field, to describe either photon absorption
or emission, one has to evaluate the matrix element of $H_I$
between the one photon state $| \vec{\epsilon} , ~ \vec{q}
\rangle$ with energy $E_\gamma = | \vec{q} |$ and polarization
$\vec{\epsilon}$ and the electromagnetic vacuum. The decay rate of
the classical nonlinear mode is controlled by the following matrix
element, computed in the limit of long wavelengths,
\begin{equation}
\langle \vec{\epsilon} , ~ \vec{q}  | H_I |  0 \rangle  =  2 \, \sqrt{\frac{2 \pi c}{q} } \, e_{sol} \, k \,
\int \, dx \,dy ~ \phi(x) \phi^\prime(x) ~
  \left( \vec{\epsilon} \cdot \hat{e}_y \right) \, ,
 \label{elemento_matriz}
\end{equation}
up to a volume normalization associated with the photon state.
In (\ref{elemento_matriz}) $c$ stands for the velocity of light in the vacuum and $\hat{e}_y$
is the unit vector along $y$ direction.

Equation (\ref{elemento_matriz}) shows that only electromagnetic waves polarized along the $y$ direction,
 i.e. for a photon polarization along the shortest direction of the graphene sheet,
are able to couple to the solitonic modes.
For nanotubes, $\hat{e}_y$ is replaced by $\hat{e}_\theta$, the unit vector along the $\theta$ direction,
 and the photon polarization needs an
$\hat{e}_\theta$ component in order to couple to the solitonic modes.

For the kink solution (section \ref{sec_kink}) the integrand function in (\ref{elemento_matriz})
is antisymmetric and the matrix element vanish. In the long wavelength limit, the kink solutions do not couple
to the photon.

For the non-linear solution of the equations of motion discussed in section \ref{sec_loehe}, it follows that
\begin{eqnarray}
\langle \vec{\epsilon} , ~ \vec{q}  | H_I |  0 \rangle
   & = &  2 \, \sqrt{\frac{2 \pi c}{q} } \, e_{sol} \, k \, \phi^2_0 \, \left( \vec{\epsilon} \cdot \hat{e}_y \right) \, W \, .
 \label{elemento_matriz_loehe}
\end{eqnarray}
It follows that the decay rate for photon production grows with the width squared and $\phi^4_0$,
i.e. the fermionic mass gap squared.
For the configuration (\ref{lohe}), it follows that, see equation (\ref{Eq_gap_E}),
\begin{eqnarray}
   \left| \langle \vec{\epsilon} , ~ \vec{q}  | H_I |  0 \rangle \right|^2  ~ = ~
  2 \, \pi \, c \, e^2_{sol} \, \frac{k^2}{q} \, \left( \vec{\epsilon} \cdot \hat{e}_y \right)^2 \,
  \frac{W^2 \, E^2_g}{g^2_2 + h^2_2}
 \label{elemento_matriz_loehe}
\end{eqnarray}
A non-vanishing matrix element requires not only a photon polarized along the shortest direction
of the graphene sheet but also a solitonic wave which transport momenta $k$ along the $y$ direction.

In what concerns the scaling of the transition rate $\Gamma$ with $W$, combining equations (\ref{eq_e_width}) and
(\ref{elemento_matriz_loehe}), it follows that for $W > 16$ nm
\begin{equation}
  \Gamma \propto   \left( \frac{W}{W - W^*} \right)^4 \, .
  \label{cross_section}
\end{equation}

%===========================================================================================
%===========================================================================================
\section{Conclusions \label{fim}}

The gauge model for graphene suggested in \cite{Jackiw_Pi_2007} and extended in \cite{Oliveira2011},
combined with the phenomenological observation of how the mass gap changes with the graphene width,
predicts that solitonic
waves can couple to the electromagnetic field, provided the photon has the right polarization.
The decay rate is non-vanishing only when the photon polarization has
a component along the $y$ direction for graphene or along the $\theta$ direction for a nanotube.

From the point of view of the mass gap, the non-linear waves can be thought as giving rise to regions in graphene
where the electrons behave as massless particles, those where $\phi \sim 0$, and regions where the electrons acquire a
finite mass, regions where $\phi \sim \phi_0$. If experimentally one can distinguish such regions by means of a spectroscopic
analysis like, for example, scanning tunneling microscope as performed in \cite{Kobayashi2005} for the different graphene terminations
or in \cite{Lahiri2010} to investigate one dimensional defects, then one could observe directly the mass profiles and
extract information on the solitonic waves, provided that the life time of these non-linear states is long enough. From these
profiles one can constraint the possible forms of the scalar potential $V( \phi )$ and, in this way, help to better define the
effective model.

The problem discussed here has several energy scales which behave differently with $W$. This different scaling
can, in principle, be explored to identify the soliton modes of section \ref{sec_loehe}. As discussed above,
$E_g \propto \phi^2_0$. Therefore, the energy associated with the soliton scales as
\begin{equation}
   E_{sol} ~ \propto ~ E^{3/2}_g \, W \quad \mbox{ and } \quad E^2_g \, W
\end{equation}
for the kink solution  (\ref{kink}) and for (\ref{lohe}), respectively. This result suggests that the energy
of the solitonic modes scales with mass gap $E_g$ to a given power and the nanoribbon width.

This different behavior can be explored, in a systematic investigation of the spectra of photon emission
with $W$, to identify the nonlinear modes in graphene nanoribbons and improve our understanding of the
effective theory discussed here.

In what concerns the typical energy associated with the solitonic
modes, the estimates performed at the end of section
\ref{energy_estimation}  suggests, for example, $E_{sol} \sim 7$
meV for a $W \sim 10$ nm for the case of an effective potential
which goes as $\varphi^6$. Then, the possible transitions
involving the nonlinear modes open a window to explore the far
infrared and terahertz frequencies. Furthermore, the coupling of
the nonlinear waves with the electromagnetic field are sensible to
the photon polarization. The nanoribbon is transparent to photons
polarized along the largest length, but not to those polarized
along the nanoribbon width. These characteristics are of interest
for optical devices as, for example, broadband polarizers applied
to photonics circuits for telecommunications. Note, however, that
similar properties can occur in association with electronic
intraband transitions as described in \cite{Hipolito2012}.
A possible way to distinguish between the two mechanism is to
investigate the behavior with the chemical potential. For  the
solitons, the results described here are independent of the
chemical potential. Dichroism with intraband transitions requires
a non-vanishing chemical potential.

We believe that our discussion can motivate experiments that could
disentangle and identify the role of the carbon structure via the presence of nonlinear waves
and/or intraband electronic transitions in grapheme nanoribbons.

%==========================================================
%==========================================================
\section*{Acknowledgements}

We thank A. J. Chaves for useful discussions and comments.
The authors acknowledge financial support from the Brazilian
agencies FAPESP (Funda\c c\~ao de Amparo \`a Pesquisa do Estado de
S\~ao Paulo) and CNPq (Conselho Nacional de Desenvolvimento
Cient\'ifico e Tecnol\'ogico). O. Oliveira acknowledges financial support from FCT under contracts
PTDC/\-FIS/\-100968/\-2008
 under the initiative QREN financed by the UE/FEDER through the Programme
COMPETE.

%==========================================================
%==========================================================

\end{document}